\documentclass[12pt,cite,epsf,epsfig]{article}
\usepackage{epsfig}
\title{Constraints on the Neutrino Parameters from the `Rise-up' in the Boron
Neutrino Spectrum at Low Energies}
\author{S. Dev\thanks{E--mail : dev5703@yahoo.com}
 and Sanjeev Kumar\thanks{E--mail : sanjeev3kumar@yahoo.co.in }\\
	{\em Department of Physics, Himachal Pradesh University,} \\
	 {\em Shimla, India-171005.}}

\begin{document}
\maketitle

\begin{abstract}
The rise-up in boron neutrino spectrum at low energies has been studied
within the framework of `pure LMA' scenario. Indirect bounds on the spectral
`upturn' have been obtained from the available solar neutrino data. These
bounds have been used to demonstrate the efficacy of the precision
measurements of the `upturn' for further constraining the neutrino parameter
space allowed by SNO salt phase data. The sterile neutrino flux has been
constrained in the light of the recent 766.3 Ty KamLAND spectral data.
\end{abstract}

Neutrino Physics is passing through a phase of spectacular development. Vast
amount of solar and atmospheric neutrino data has been accumulated and the
neutrino deficits have been established to be the consequence of
non-standard neutrino physics. The most recent steps in this direction are
the pioneering results from SNO and KamLAND experiments. The SNO experiment
provided a model independent proof of solar neutrino oscillations and the
terrestrial disappearance of reactor $\overline{\nu }_{e}$ in the KamLAND\
experiment has provided a further confirmation of the neutrino oscillation
solution of the solar neutrino problem (SNP). This gives us confidence in
the oscillation solution of the atmospheric neutrino anomaly.

The neutral current measurements at SNO \cite{1} have, conclusively,
established the oscillations of solar neutrinos. After the evidence of
terrestrial antineutrino disappearance in a beam of electron antineutrinos
reported by KamLAND \cite{2}, all other \cite{3} explanations of the solar
neutrino deficit can, at best, be just subdominant effects. After these
pioneering experiments, there is no scope for doubting the physical reality
of neutrino mass and the consequent oscillations. KamLAND\ is the first
experiment to explore the neutrino parameter space relevant to SNP with a
beam of terrestrial neutrinos and has, convincingly, demonstrated the
existence of neutrino oscillations confined to the large mixing angle (LMA)
region. The total event rate as well as the spectrum distortion at KamLAND
are in good agreement with the LMA expectations. Recently, updated analyses
of all the available solar and reactor neutrino data including KamLAND\ and
SNO salt phase data have been presented \cite{4}. However, even after the
confirmation of the LMA MSW mechanism as a dominant solution of SNP, the
oscillation parameters are not precisely known. A precise determination of
these parameters will be of great importance for theory as well as
phenomenology of neutrino oscillations in particular and particle physics in
general.

The solar neutrino experiments have, already, entered a phase of precision
measurements for oscillation parameters. On the other hand, the LMA solution
is facing a deeper scrutiny. In fact, the completeness of the LMA solution
is being questioned \cite{5} and the scope for some possible subdominant
transitions is being explored \cite{6,7} vigorously. Does the LMA solution
satisfactorily explain all the solar neutrino data? Are there any
observations indicating new physics beyond LMA? These are some of the
relevant questions being posed. It is also the high time to put the LMA
predictions to closer experimental scrutiny. There are, at least, two
generic predictions of LMA \cite{6} which point towards life beyond LMA. One
of these is the prediction of a high argon production rate, $Q_{Ar}\approx
3SNU$, for the Homestake experiment which is about $2\sigma $ above the
observed rate. Another generic prediction of the LMA scenario is the
`spectral upturn' at low energies. Within the LMA\ parameter space, the
survival probability should increase with decrease in energy and for the
best fit point, the upturn could be as large as 10-15\% between 8MeV and
5MeV \cite{6}. However, neither the SuperKamiokande (SK) nor SNO have
reported any statistically significant `rise-up' in the observed neutrino
survival probability. Both these predictions of LMA can only be tested in
the forthcoming phase of high precision measurements in the solar neutrino
experiments and are crucial for confirmation of the LMA solution.

The distortions in the neutrino spectrum are an important factor in
resolving the solar neutrino problem. These distortions arise due to the
energy dependence of the survival probability as a result of which neutrinos
with different energies survive in different proportions leading to
distortions in the observed spectrum. Experimentally, the boron neutrinos
are the most accessible source for the study of the distortions in the
observed spectrum since the SK and SNO detect the boron neutrinos in the
small energy bins over a wide energy range. Since, the LMA has emerged as a
solution of the SNP, the spectrum distortions within the LMA scenario are of
paramount importance for the final confirmation of the LMA as a solution of
the SNP and, also, for possible physics beyond LMA.

In the present work, we focus on the `rise-up' in the neutrino spectrum at
low energies and demonstrate how a precision measurement of the `upturn' can
be used to further constrain the neutrino parameter space allowed by the SNO
salt phase data. In the absence of concrete experimental results on the
`rise-up', we obtain indirect bounds on the `rise-up' in the boron neutrino
spectrum by comparing the boron neutrino survival probability obtained from
the experiments with the asymptotic value of the corresponding LMA survival
probability.

The apparent lack of the `rise-up' in the observed boron neutrino spectrum
at low energies has been sought to be explained by introducing subdominant
transitions into sterile neutrinos \cite{6} and/or antineutrinos \cite{7}.
In the present work, it has been shown that the `rise-up' in the boron
neutrino spectrum can be reduced significantly by choosing a suitable point
within the LMA parameter space itself. We examine the status of this
`rise-up' within the pure LMA scenario in the following manner. From the SNO
salt phase data and the value of the neutrino mixing angle `$\theta $'
obtained from the global analyses, indirect bounds on the `rise-up' are
obtained. The constraints on the `rise-up' and the boron neutrino survival
probability are combined to further constrain the neutrino parameter space
allowed by the SNO.

The LMA survival probability \cite{8}, to a very good approximation, can be
written as 
\begin{equation}
P=\frac{1}{2}+\frac{1}{2}\cos 2\theta \cos 2\theta _{m},
\end{equation}
where the mixing angle in matter is given by 
\begin{equation}
\cos 2\theta _{m}=\frac{\cos 2\theta -\beta }{\sqrt{\left( \cos 2\theta
-\beta \right) ^{2}+\sin ^{2}2\theta }}
\end{equation}
and the ratio of matter to vacuum effects `$\beta $' is given by 
\begin{equation}
\beta =\frac{2\sqrt{2}G_{F}N_{e}E}{\Delta m^{2}}.
\end{equation}
E is the energy of the neutrino and $N_{e}$ is the electron number density
at the point of maximal boron neutrino production i.e.e at. $x=r/R_{S}=0.05$
where $R_{S}$ is the solar radius so that 
\begin{equation}
G_{F}N_{e}=0.4714\times 10^{-11}eV
\end{equation}
at this point \cite{9}. The energy dependence of the LMA survival
probability P given by eqn. (1) is shown in Fig.1 (dashed line) along with
its asymptotic value $sin^{2}\theta $ (dotted line). The survival
probability averaged over the production region of the boron neutrinos \cite
{9} has been plotted as a solid line. It can be easily seen that the
analytical expression (1) is in fairly good agreement with the exact
numerical result. The value of P is slightly increased by averaging over the
production region. Moreover, the earth regeneration effects will, also,
increase the survival probability only by a small amount. It can be seen
that the percentage increase in the survival probability from the earth
regeneration effects equals the day-night asymmetry. The expected day-night
asymmetry at SNO is about 3\% \cite{10}. Thus, eqn. (1) is a fairly good
approximation to survival probability.
\begin{figure}[ht]
\begin{center}
\rotatebox{270}{\epsfig{file=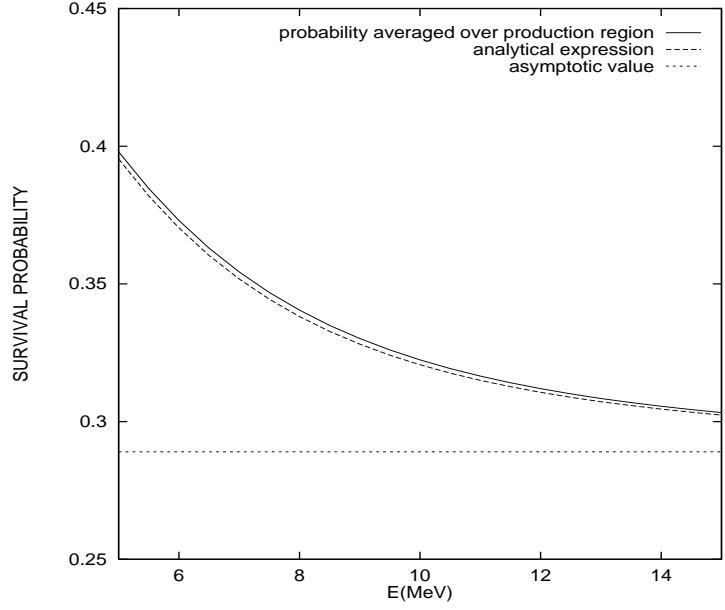, width=8.0cm, height=10.0cm}}
\end{center}
\caption{The energy dependence of boron neutrino survival probability in the
LMA scenario for $\Delta m^{2}=7.1\times 10^{-5}eV^{2}$ and $\theta =32.5$
degrees.}
\end{figure}

Equation (1) can be written as 
\begin{equation}
P=sin^{2}\theta +R,
\end{equation}
where 
\begin{equation}
R=\cos 2\theta \cos ^{2}\theta _{m}
\end{equation}
is the rise-up in the survival probability. Obviously, R is always positive
and increases with decrease in energy. The survival probability P is an
increasing function of both $\Delta m^{2}$ and $\theta $ in the allowed LMA
region in contrast to R which is an increasing function of $\Delta m^{2}$
and a decreasing function of $\theta $, within this region. The `rise-up' R
becomes zero for maximal mixing. Since, maximal mixing is rejected at 5.4
standard deviations, the `rise-up' cannot be zero. Hence, a non-zero
`rise-up' is an inescapable consequence of the LMA scenario.

Global analysis of the SNO salt phase data along with other solar and
reactor neutrino data yields \cite{11} 
\begin{equation}
\Delta m^{2}=7.1_{-0.6}^{+1.2}\times 10^{-5}eV^{2},
\end{equation}
\begin{equation}
\theta =32.5_{-2.3}^{+2.4}\deg .
\end{equation}
For these LMA parameters, we have 
\begin{equation}
\sin ^{2}\theta =0.289_{-0.036}^{+0.038},
\end{equation}
\begin{equation}
P=0.362_{-0.031}^{+0.036},
\end{equation}
\begin{equation}
R=0.074_{-0.025}^{+0.044}.
\end{equation}
It is clear that R is about three standard deviations above zero and is
large enough to be measured experimentally.

The value of the survival probability for the boron neutrinos can be
calculated from the SNO CC and NC rates using the relation

\begin{equation}
P=\frac{\phi _{CC}^{SNO}}{\phi _{NC}^{SNO}}
\end{equation}
where we have assumed transitions into active flavors only. Transitions into
sterile neutrinos can be important and will be studied elsewhere \cite{12}.
Even though, neither SK nor SNO has reported any statistically significant
`rise-up', one can infer the `rise-up' at 6.4MeV from SNO CC/NC ratio.
Since, $sin^{2}\theta $ is constrained by equation (9), we can constrain R
using eqns. (5) and (9). In this manner, we can obtain an indirect upper
bound on R. However, the value of $\theta $ obtained from the global
analyses is not model independent as a result of which the value of R
obtained in this manner will be model dependent and will be valid only
within the LMA scenario.

The pure $D_{2}O$ data from SNO \cite{1} gives 
\begin{equation}
\phi _{CC}^{SNO}=1.76_{-0.103}^{+0.108}\times 10^{6}cm^{-2}s^{-1},
\end{equation}
\begin{equation}
\phi _{NC}^{SNO}=6.42_{-1.67}^{+1.66}\times 10^{6}cm^{-2}s^{-1},
\end{equation}
where the statistical and the systematic errors have been combined in
quadratures.. From equation (12), we have 
\begin{equation}
P=0.274_{-0.073}^{+0.073}.
\end{equation}
Using the LMA value of $\theta $ and equation (5), one can obtain 
\begin{equation}
R=-0.015_{-0.081}^{+0.082},
\end{equation}
which is not, significantly, different from zero. However, one can obtain an
upper bound on R from equation (16) viz. 
\begin{equation}
R\leq 0.120
\end{equation}
at 90\%C.L. It may be worthwhile to mention that the NC rate given in
equation (14) has been obtained without any assumptions regarding the energy
dependence of the survival probability. If one assumes an undistorted boron
neutrino spectrum and, hence, an energy independent survival probability,
SNO pure $D_{2}O$ data gives 
\begin{equation}
\phi _{NC}^{SNO}=5.09_{-0.608}^{+0.637}\times 10^{6}cm^{-2}s^{-1}.
\end{equation}
Using this value instead of the value quoted in equation (14) would give 
\begin{equation}
P=0.346_{-0.046}^{+0.048}
\end{equation}
and 
\begin{equation}
R=0.057_{-0.058}^{+0.061}
\end{equation}
in agreement with the LMA values given in eqns. (10) and (11). However, the
LMA survival probability being energy dependent, the use of the value quoted
in equation (18) for deriving constraints on neutrino parameters will not be
internally consistent \cite{13}.

The most recent SNO salt phase data \cite{11} 
\begin{equation}
\frac{\phi _{CC}^{SNO}}{\phi _{NC}^{SNO}}=0.306_{-0.035}^{+0.035}
\end{equation}
can, also, be used to obtain the new bounds on P and R viz. 
\begin{equation}
P=0.306_{-0.035}^{+0.035},
\end{equation}
\begin{equation}
R=0.017_{-0.050}^{+0.052}.
\end{equation}
This value of `rise-up' will be used henceforth. The value of P given in
equation (22) is smaller than the mean LMA value by an amount 
\begin{equation}
0.057_{-0.056}^{+0.068}
\end{equation}
which is one standard deviation above zero. We shall explore the allowed LMA
region to reduce the difference between the LMA values of P, R and their
experimental values given by eqns. (22) and (23) respectively which imply
the following upper bounds on P and R: 
\begin{equation}
P\leq 0.363,
\end{equation}
\begin{equation}
R\leq 0.102,
\end{equation}
at 90\%C.L. As noted earlier, the `rise-up' R becomes smaller for smaller
values of $\Delta m^{2}$ and larger values of $\theta $. However, a larger
value of $\theta $ leads to an increase in the value of P. In fact, the
experimental value of P is already greater than the mean LMA value and
cannot be increased further. Hence, we consider the constraints (25) and
(26) on P and R simultaneously. This can be achieved by plotting the
constant P and constant R curves in the allowed parameter space. The curves
corresponding to 90\% C.L. upper bounds on P and R have been plotted in
Fig.2 within the LMA parameter space allowed by the SNO. The overlap region
below P and R curves is the region of parameter space allowed by the bounds
on `rise-up' and survival probability obtained above. The resulting upper
bounds on $\Delta m^{2}$ and $\theta $ are 
\begin{equation}
\Delta m^{2}\leq 7.9\times 10^{-5}eV^{2},
\end{equation}
\begin{equation}
\theta \leq 33.7\deg ,
\end{equation}
at 90\% C.L. Thus, the `rise-up' in the boron neutrino spectrum can be used
to further restrict the neutrino parameter space. In fact, the bound on the
`rise-up' derived from SNO salt-phase data selects lower values of $\Delta
m^{2}$ consistent with the conclusions reached by Aliani et al \cite{4} who
incorporated the SNO spectrum data in the global analysis. Therefore, the
`pure LMA' scenario will get rejected at more than 90\% C.L. if the future
precision measurements favor $\Delta m^{2}>7.9\times 10^{-5}eV^{2}$. The
value of $\Delta m^{2}$ larger than $7.9\times 10^{-5}eV^{2}$ will be clear
signature of physics beyond LMA being manifest in the oscillations of solar
boron neutrinos. The inclusion of the earth regeneration effect as well as
the averaging over the production region will only decreases the value of $%
\Delta m^{2}$ and the upper bound mentioned above will, still, remain
valid.


\begin{figure}[ht]
\begin{center}
\rotatebox{270}{\epsfig{file=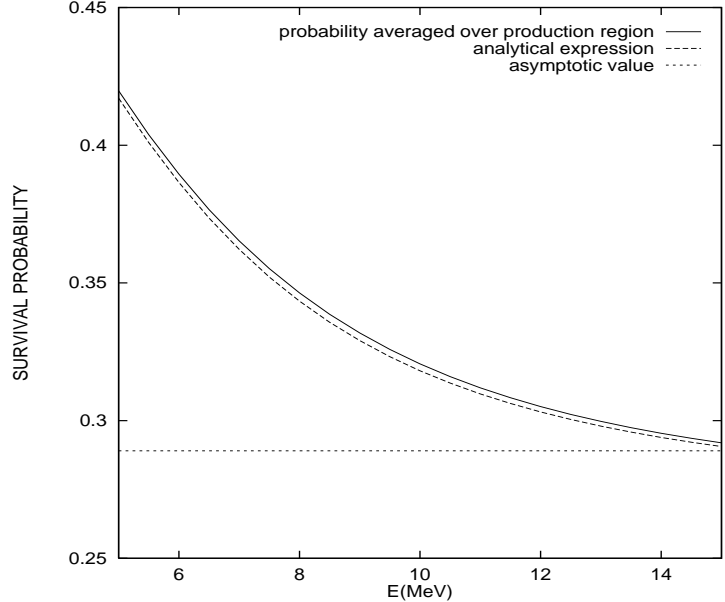, width=8.0cm, height=10.0cm}}
\end{center}
\caption{The constant P and the constant R curves plotted within the neutrino
parameter space allowed by SNO salt phase data.}
\end{figure}
The curves P=0.342 and R=0.070 corresponding to 1.02$\sigma $C.L. are also
shown in Fig. 2 below which there is no overlap. These two curves intersect
at 
\begin{equation}
\Delta m^{2}=6.5\times 10^{-5}eV^{2},
\end{equation}
\begin{equation}
\theta =31.4\deg .
\end{equation}
For these values of $\Delta m^{2}$ and $\theta $, the difference between the
LMA values of P, R and their experimental values (22) and (23) is the least
(about one standard deviation). This can be regarded as the best fit point
in the SNO allowed parameter space. The values of $\Delta m^{2}$ and $\theta 
$ obtained from the global analyses of all the solar neutrino data \cite{10}
are very close to the values obtained here.

While this work was in progress, KamLAND reported 766.3 Ty spectrum data 
\cite{14} which has been combined with the solar neutrino data by several
authors \cite{4,10,14}. The two main implications of the new KamLAND data
are the increase in the value of $\Delta m^{2}$ to $8.3_{-0.37}^{+0.40}%
\times 10^{-5}eV^{2}$ and a decrease in the best-fit value of $\theta $ to $%
31.3_{-1.3}^{+1.9}\deg $ \cite{10}. The best fit value of $\Delta m^{2}$
obtained in \cite{10} is larger than the upper bound derived here ( eqn.
(27) ) hinting towards possible new physics beyond LMA.

The inclusion of the earth regeneration effects will increase the values of
P and R by only about 3\% which is too small as compared to the rise-up
(which is about 28\%) [ see Fig. 3]. Moreover, the LMA values of P and R
are, already, larger than their experimental values. The earth regeneration
effect will, therefore, further increase their values enhancing the mismatch
between the theory and experiment. This would make the upper bound on $%
\Delta m^{2}$ even more restrictive.
\begin{figure}[ht]
\begin{center}
\rotatebox{270}{\epsfig{file=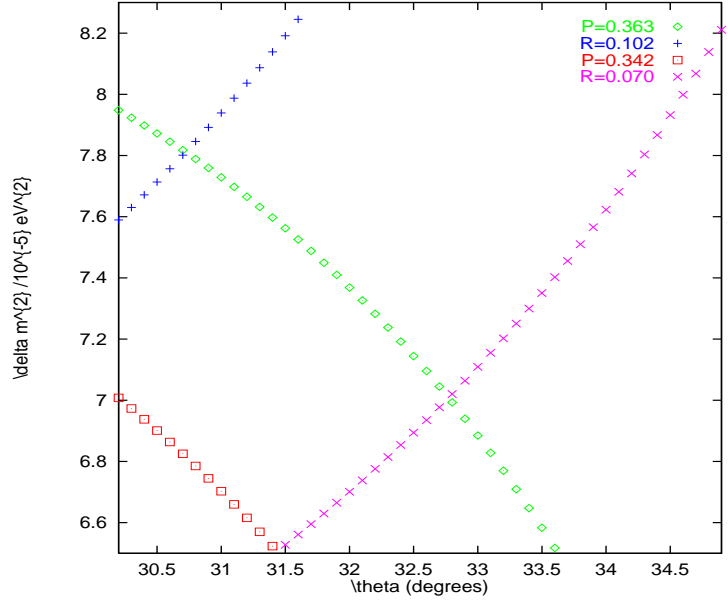, width=8.0cm, height=10.0cm}}
\end{center}
\caption{The same as Fig.1 for the values of $\Delta m^{2}=8.3\times
10^{-5}eV^{2}$ and $\theta =31.3$ degrees.}
\end{figure}

For these values of $\Delta m^{2}$ and $\theta $, P and R will, now, become 
\begin{equation}
P=0.376_{-0.014}^{+0.017},
\end{equation}
\begin{equation}
R=0.106_{-0.025}^{+0.023},
\end{equation}
in place of eqns. (10) and (11). The difference of P from its experimental
value (eqn. (22)) is given by 
\begin{equation}
0.076_{-0.038}^{+0.039}
\end{equation}
which is $2\sigma $ above zero. Hence, there is considerable difference
between the experimental and theoretical values of P in the LMA scenario and
we are constrained to go beyond the pure LMA scenario. A natural candidate
for these transitions would be the spin flavor precession (SFP) driven
transitions into antineutrinos. Since, there are very stringent bounds on
the solar antineutrino flux \cite{15}, the transitions into antineutrinos
can not account for this difference. Hence, we attribute the whole of this
difference to the transitions into sterile neutrinos and obtain \cite{16} 
\begin{equation}
P(\nu _{e}\rightarrow \nu _{e})=\frac{x\sin ^{2}\alpha }{1-x\cos ^{2}\alpha }%
,
\end{equation}
\begin{equation}
P(\nu _{e}\rightarrow \nu _{\mu })=\left( 1-P(\nu _{e}\rightarrow \nu
_{e})\right) \sin ^{2}\alpha ,
\end{equation}
\begin{equation}
P(\nu _{e}\rightarrow \nu _{S})=\left( 1-P(\nu _{e}\rightarrow \nu
_{e})\right) \cos ^{2}\alpha ,
\end{equation}
where 
\begin{equation}
x=\frac{\phi _{CC}^{SNO}}{\phi _{NC}^{SNO}},
\end{equation}
and 
\begin{equation}
P(\nu _{e}\rightarrow \nu _{\mu })=1-P_{LMA}.
\end{equation}
Here, $\alpha $ is the sterile mixing angle. From these equations, we obtain 
\begin{equation}
P(\nu _{e}\rightarrow \nu _{e})=\frac{x\left( 1-P_{LMA}\right) }{\left(
1-x\right) },
\end{equation}
and 
\begin{equation}
\sin ^{2}\alpha =1-\frac{P_{LMA}-x}{1-2x+xP_{LMA}}.
\end{equation}
Using equation (21) for $x$ and equation (31) for $P_{LMA}$, we obtain 
\begin{equation}
\sin ^{2}\alpha =0\allowbreak .\,861_{-0.077}^{+0.091},
\end{equation}
and

\begin{equation}
P(\nu _{e}\rightarrow \nu _{e})=0.\,275_{-0.049}^{+0.055}.
\end{equation}
The pure sterile solution ($\sin ^{2}\alpha =0$) is disfavored at 11.2
standard deviations. From equation (36), we obtain 
\begin{equation}
P\left( \nu _{e}\rightarrow \nu _{s}\right) =0\allowbreak
.101_{-0.069}^{+0.066},
\end{equation}
at 3$\sigma $C.L$.$which implies 
\begin{equation}
P\left( \nu _{e}\rightarrow \nu _{s}\right) \leq 0.299.
\end{equation}
The sterile flux is non-zero at about 1.5 standard deviations. A more
elaborate analysis is needed to constrain the sterile component using the
approach adopted here and will be presented elsewhere \cite{12}.

In conclusion, the `rise-up' in the boron neutrino spectrum at low energies
has been studied within the framework of the LMA scenario. Indirect bounds
on the rise-up have been obtained from the available solar neutrino data.
These bounds have been used to demonstrate as to how a precision measurement
of the rise-up can be used to further constrain the neutrino parameter space
allowed by the SNO salt phase data. It is found that the pure LMA solution
is sufficient to explain the SNO salt phase data for $\Delta m^{2}\leq
7.9\times 10^{-5}eV^{2}$ and $\theta \leq 33.7\deg $ since larger values of $%
\Delta m^{2}$ will violate the upper bound given in equation (26). However,
the most recent global analyses \cite{4,10,14} of the solar neutrino and the
recent KamLAND data favor a value of $\Delta m^{2}$ which violates this
upper bound. Consequently, pure LMA solution seems to be disfavored and
other subdominant transitions seem unavoidable. The theoretical and
experimental values of the boron neutrino survival probability in the pure
LMA scenario for the most recent LMA parameters differ by two standard
deviations. This discrepancy is too large to be explained by the subdominant
SFP transitions into antineutrinos. In the present work, this discrepancy
has been attributed to the subdominant transitions into the sterile
neutrinos. It is concluded that the sterile neutrino flux in this scenario
could be as large as 0.299 times the boron neutrino flux at $3\sigma $.

{\it Acknowledgments}: One of the authors (SK) gratefully acknowledges the
financial support provided by the Council for Scientific and Industrial
Research (CSIR), Government of India.


\end{document}